\documentclass[twocolumn]{aastex631}

\usepackage{color}
\usepackage{natbib}
\usepackage{graphicx}
\usepackage{ulem}
\usepackage{wrapfig}
\usepackage{amsmath}

\newcommand{\um}{$\mu$m}
\newcommand{\htp}{H$_3^+$}

\received{\today}

\shorttitle{No H$_3^+$ Emission Seen in the Hot Jupiters WASP-80b and WASP-69b}
\shortauthors{Richey-Yowell et al. (2025)}

\begin{document}

\title{Stringent Limits on H$_3^+$ Emission from the Hot Jupiters WASP-80b and WASP-69b}

\author[0000-0003-1290-3621]{Tyler Richey-Yowell}
\affil{Lowell Observatory, 1400 W. Mars Hill Road, Flagstaff, AZ 86001, USA}
\affil{Percival Lowell Postdoctoral Fellow}
\email{try@lowell.edu}

\author[0000-0002-7260-5821]{Evgenya L. Shkolnik}
\affil{School of Earth and Space Exploration, Arizona State University, Tempe, AZ 85281, USA}

\author[0000-0003-4450-0368]{Joe Llama}
\affil{Lowell Observatory, 1400 W. Mars Hill Road, Flagstaff, AZ 86001, USA}

\author[0000-0002-3522-5846]{James Sikora}
\affil{Lowell Observatory, 1400 W. Mars Hill Road, Flagstaff, AZ 86001, USA}

\author[0000-0002-9946-5259]{Peter Smith}
\affil{School of Earth and Space Exploration, Arizona State University, Tempe, AZ 85281, USA}

\begin{abstract}

Observations of auroras on exoplanets would provide numerous insights into planet-star systems, including potential detections of the planetary magnetic fields, constraints on host-star wind properties, and information on the thermal structures of planets. However, there have not yet been any discoveries of auroras on exoplanets. In this paper, we focus on the search for infrared auroral emission from the molecular ion \htp\,, which is common in the atmospheres of  solar system planets Jupiter, Saturn, and Uranus. Using Keck/NIRSPEC high-resolution spectroscopy, we search for \htp\, emission from two hot Jupiters, WASP-80b and WASP-69b. We do not see any evidence of emission in the observed spectra when cross-correlating with an \htp\, spectral model or when using an auto-correlation approach to search for any significant features. We therefore place upper limits on the total emission of $5.32 \times 10^{18}$ W for WASP-80b and $1.64 \times 10^{19}$ W for WASP-69b. These upper limits represent the most stringent limits to date and approach the regime of emission suspected from theoretical models. 

\end{abstract}


\section{Introduction}\label{sec:intro}

Auroras are photon emission caused by interactions between celestial bodies and external particles. Searches for auroral emission from exoplanets have been gaining popularity due to the unique science these types of detections would provide. Auroras from exoplanets can provide a detection and potential measurement of the planetary magnetic field, probe the stellar wind environment, and give information about the thermal profile of the planet atmosphere \citep[e.g.][]{Miller2000}. 

Two main methods for discovery are being explored: one through radio emission due to electron-cyclotron maser instability (ECMI) emission from the travel of the particle along the magnetic field of the planetary body, and the other through molecular emission due to the interaction between the electron and the planetary atmosphere \citep[see review by][]{Callingham2024NatAs...8.1359C}. There have been a few suggestions of auroral interactions observed in the radio with LOFAR \citep{Turner2021A&A...645A..59T, Vedantham2020NatAs...4..577V}, but yet no confirmed detection. One of the reasons for these lack of detections could be that the ECMI emission cannot propagate in atmopsheres of hot Jupiters, which have been the main target of studies.  \citet{Weber2017} suggest that the extended ionospheres of these hot Jupiters cause too large of plasma densities in their magnetospheres and thus ECMI is not efficient. Therefore, molecular emission may then be the preferred method for finding these auroras on hot Jupiters.

The auroral emission from the molecular ion H$_3^+$ is the dominant cooling mechanism in Jupiter's thermosphere and is a primary probe of temperature and ion densities \citep{Miller2000}. Dubbed as ``the H$_3^+$ thermostat", this emission was first detected in Jupiter by \citet{Drossart1989} and then subsequently in Saturn \citep{Geballe1993} and Uranus \citep{Trafton1993}. In Jupiter, the magnetically-fueled H$_3^+$ auroras at the north and south poles are $\sim$100 times brighter than the disk emission \citep{O'Donoghue2016}. This is achieved by collisions with energetic electrons funneled down magnetic field lines and stellar extreme ultraviolet (EUV) flux ionizing H$_2$, with H$_2^+$ then interacting with a neutral H$_2$:

\begin{equation}
\begin{split}
    &\text{H$_2$} + e^* \rightarrow \text{H}_2^+ + e + e \\
    &\quad\text{H}_2 + h\nu \rightarrow \text{H}_2^+ + e \\
    &\;\;\;\text{H}_2 + \text{H}_2^+ \rightarrow \text{H}_3^+ + \text{H}
\end{split}
\end{equation}

Jupiter emits 10$^{12}$ W in the emission lines of H$_3^+$ \citep{Lam1997}. At a distance of 10 pc, the resulting flux of 8 x 10$^{-25}$ W/m$^2$ would be undetectable. A hot Jupiter ($a$ $\approx$ 0.05 AU) is 100 times closer to its parent star and experiences at least 10$^4$ times the EUV flux and magnetic interaction with the stellar magnetosphere \citep{Shkolnik2005, Shkolnik2008}. Direct detection of a transiting or even non-transiting hot Jupiter atmosphere is possible because a significant fraction of this additional energy is re-radiated by narrow lines of molecular coolants, the strongest of which is H$_3^+$. 

There is some debate in the literature about the expected levels of H$_3^+$ emission from hot Jupiters, ranging from $\geq$10$^{17}$ W \citep{Miller2000} to $\sim$10$^{16}$ W \citep{Yelle2004, Chadney2016A&A...587A..87C} to $\sim$10$^{15}$ W \citep{Koskinen2007}. However, in all cases, the models do not include the possibility of a planetary magnetic field. A planetary field would trap ions, limiting ion escape, and cause the precipitation of electrons and ions along magnetic field lines, producing polar enhancements of H$_3^+$ on the close-in planets similar to those observed on our Jupiter. This might increase the H$_3^+$ emission from a hot Jupiter by orders of magnitude beyond the predictions. If detected, there will be no ambiguity as to the origin of the signature as H$_3^+$ does not form in stellar atmospheres.

H$_3^+$ has a strong ro-vibrational spectrum emitting strongest between 3 -- 4 $\mu$m, which has been the focus of most exoplanet \htp\, searches thus far. \citet{Shkolnik2006} searched for H$_3^+$ from twelve planets orbiting five F and G type stars plus one M dwarf, using the single-order spectrograph CSHELL at the IRTF. The authors focused their search on the emission from the Q(1,0) transition of H$_3^+$ at 3.953 $\mu$m. \citet{Lenz2016} used CRIRES at the VLT to search for the Q(1,0) and the 3.985 $\mu$m Q(3,0) transitions of H$_3^+$ from HD 209458b. The authors utilized both direct observations and auto-correlation between the nights, but again did not detect any signatures. \citet{Gibbs2022} conducted a search with Keck/NIRSPEC in the KL band, focusing on 3.94 -- 4.02 \um region for auroral \htp\, emission from eleven giant exoplanets around five FGKM stars. They focused on direct observations of several lines including the Q(1,0) and the Q(3,0) transitions as well. Again, no auroral emission was detected; however, they did place the most stringent limits to date of 2.2 x $10^{17}$ W for the Q(1,0) transition emission from GJ 876c and 5.2 x $10^{17}$ W for the Q(3,0) transition from GJ 876b. 

In this paper, we search for \htp\, molecular emission from two hot Jupiters that potentially represent some of the best chances for finding auroral signatures. In addition to a direct search and auto-correlation, we utilize for the first time a cross-correlation of the data with a model, an approach which has been successful for dozens of planets and molecules. We describe the two systems in Section \ref{sec:system}, after which we detail the observations and data analysis in Section \ref{sec:obs/data}. We discuss our direct, cross-correlation, and auto-correlation approaches for searching for \htp\, in Section \ref{sec:search}. Since we did not detect  emission through any of these approaches, we outline our method for determining upper limits in Section \ref{sec:upperlimits}. Finally, we discuss the implications for these non-detections in Section \ref{sec:conclusions}.

\section{System Information}\label{sec:system}

Our target strategy is to focus on the low-mass stars where the contrast of any H$_3^+$ emission will be greater than for solar-type planet hosts. These intrinsically more active stars will also likely produce greater particle and EUV/XUV flux with which to form the H$_3^+$. Our targets focus on K stars, since hot Jupiters around K stars may offer the balance between this increased radiation while not having quite enough that will instead push the molecule to higher pressures and therefore dissociate more easily \citep{Chadney2016A&A...587A..87C}. Additionally, planets with escaping helium detections at 10830\AA\, may produce more H$_3^+$ emission, as collisional interactions between the metastable helium and neutral hydrogen produce more H$^+$, which then combines with H$_2$ to form H$_3^+$ \citep{Oklopcic2018, Gonzalez-Lezana2013}. Both of our targets have detected 10830\AA\, He in their escaping atmospheres \citep{Sedaghati2017, Nortmann2018}.  

\subsection{WASP-80b}


WASP-80b is a tidally-locked $0.999^{+0.030}_{-0.031}$ $M_J$ mass \citep{Triaud2015MNRAS.450.2279T} planet orbiting a K7V star $(4145 \pm 100$ K, \citealt{Triaud2013}). The planet was discovered via primary transit and has an orbital period of $3.06785251 \pm 0.00000018$ days \citep{Kokori2023ApJS..265....4K}. The system has a velocity $V_{sys}$ of $9.821 \pm 0.768$ km/s \citep{GaiaDR2} and the planet has a radial velocity semi-amplitude $K_{p}$ of 122 $\pm$ 4 km/s, used to detect  H$_2$O, CH$_4$, NH$_3$, HCN, and tentatively CO$_2$ in the planet atmosphere via high-resolution cross-correlation by \citet{Carleo2022AJ....164..101C}. JWST NIRCam data from \citet{Bell2023} confirms methane in the atmopshere of WASP-80b. Near-infrared broad-band transmission spectroscopy by \citet{Sedaghati2017} showed evidence of He in the atmosphere. 

\subsection{WASP-69b}

WASP-69b is a $0.260 \pm 0.017 $ $M_J$ mass planet orbiting a K5V star $(4715 \pm 50$ K,  \citealt{Anderson2014}). The planet was discovered via primary transit and has an orbital period of $3.86813888 \pm 0.00000091$ days \citep{Kokori2023ApJS..265....4K}. The system has a velocity $V_{sys}$ of $-9.372 \pm 0.210$ km/s \citep{GaiaDR2} and the planet has a radial velocity semi-amplitude $K_{p}$ of  127.11$^{+1.49}_{-1.52}$ km/s, used to detect CH$_4$, NH$_3$, CO, C$_2$H$_2$, and H$_2$O in the planet atmosphere via high-resolution cross-correlation \citep{Guilluy2022A&A...665A.104G}. \citet{Schlawin2024AJ....168..104S}  observed H$_2$O, CO$_2$, and CO with JWST NIRCam and MIRI data, but did not detect any clear signatures of CH$_4$. \citet{Nortmann2018} detected an extended He atmosphere.

\section{Observations and Data Analysis}\label{sec:obs/data}

\begin{figure}[t]
    \centering
    \includegraphics[width=0.9\linewidth]{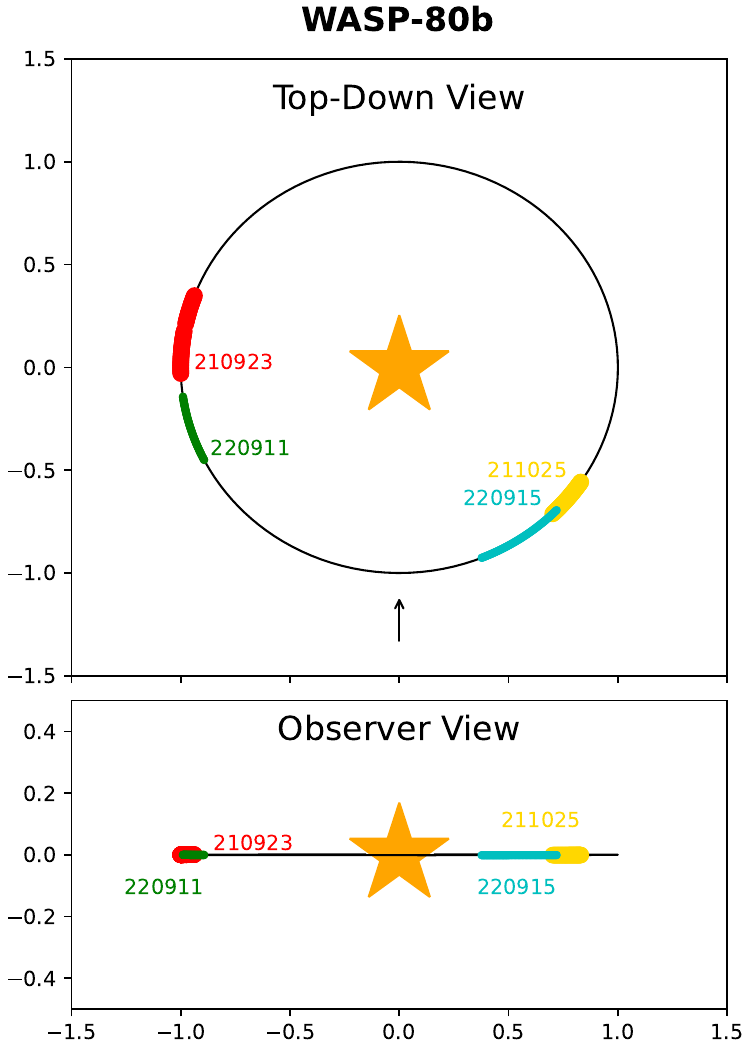}
    \singlespace\caption{Phase coverage of observations for WASP-80b. The units of the axes are arbitrary. The observation dates are in UT and written as YYMMDD. The arrow in the Top-Down View designates the line of sight from Earth.}
    \label{fig:phases80}
\end{figure}

\begin{figure}[t]
    \centering
    \includegraphics[width=0.9\linewidth]{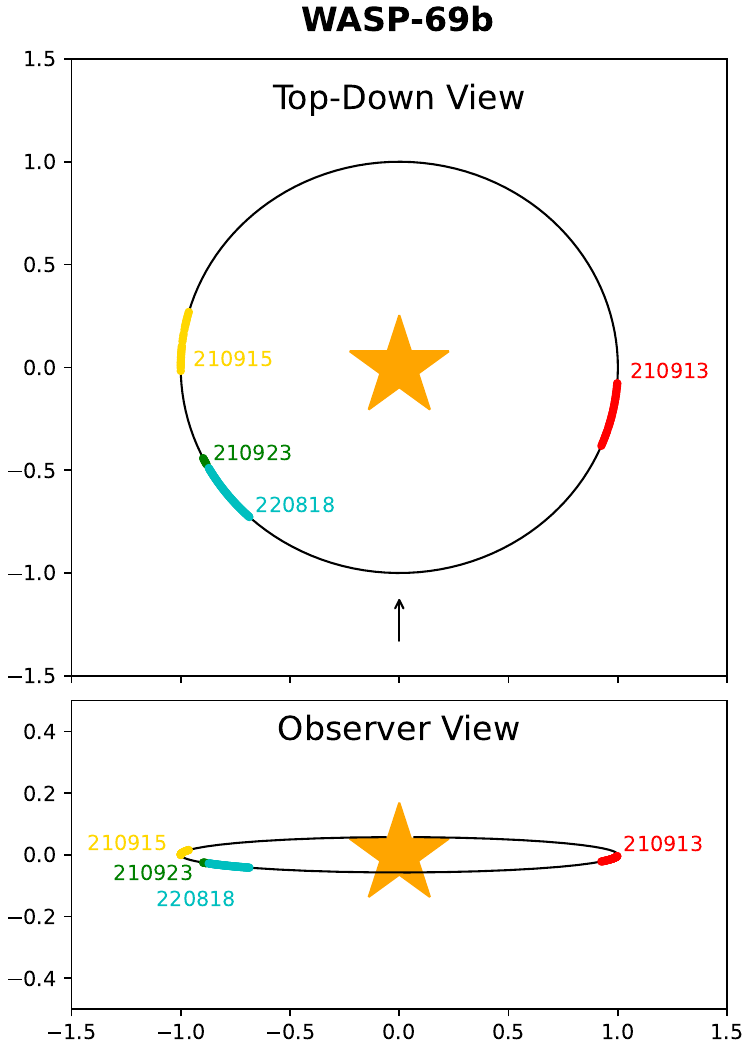}
    \singlespace\caption{Same as Figure \ref{fig:phases80} but for WASP-69b. The orbit of WASP-69b is more inclined than WASP-80b.}
    \label{fig:phases69}
\end{figure}


\begin{deluxetable*}{c c c c c}[t]
\centering
\tablecaption{\normalsize{Observing summary.}\label{tab:tested}} 
\tablehead{\colhead{Night (YYMMDD)} & \colhead{Object} & \colhead{Phase Coverage} & \colhead{\# of Spectra} & \colhead{SNR per Spectrum}}
\startdata
210923	&	WASP-80b	&	0.69 -- 0.75	&	95	&	77 -- 144	\\
211025	&	WASP-80b	&	0.12 -- 0.16	&	55	&	74 -- 163	\\
220911	&	WASP-80b	&	0.77 -- 0.82	&	87	&	71 -- 188	\\
220915	&	WASP-80b	&	0.06 -- 0.13	&	114	&	69 -- 190	\\
\hline
210913	&	WASP-69b	&	0.19 -- 0.24	&	39	&	62 -- 92	\\
210915	&	WASP-69b	&	0.71 -- 0.75	&	26	&	47 -- 87	\\
210923	&	WASP-69b	&	0.82 -- 0.83	&	12	&	113 -- 172	\\
220818	&	WASP-69b	&	0.83 -- 0.88	&	83	&	117 -- 349	\\
\enddata
\end{deluxetable*}

\subsection{Telescope, Instrument Settings, and Observational Details}
We were awarded four half-nights on Keck in both semesters 2021B and 2022B, for a total of eight half-nights on NIRSPEC/NIRSPAO \citep{NIRSPEC}. Of these, one half-night during semester 2022B was clouded out and the data were not usable. We observed from phases 0.06 -- 0.15 / 0.69 -- 0.82 for WASP-80b and 0.18 -- 0.23 / 0.70 -- 0.87 for WASP-69b (see Figures \ref{fig:phases80} and \ref{fig:phases69}) in order to balance the need for large Doppler shifts from the planets' orbits with the need for a large \textit{change in} Doppler shift necessary for to carry out the cross-correlation. The calculated velocity change of the planet and thus the \htp signature covers 4 (for the shortest observing nights and closer to quadrature) -- 12 (for the longer observing nights and farther from quadrature) pixels per night. 

Observations of Jupiter and Saturn have shown the auroral \htp\, emission to be along the entire circumference of the poles \citep[][ and references therein]{O'Donoghue2022RemS...14.6326O}; however, brighter regions of auroral ``dawn storms" have been shown to originate on the night-side of the planets, perhaps due to reconfigurations of the tails of the planetary magnetic fields \citep{Bonfond2021AGUA....200275B}. Therefore, night-side/dawn observations may have increased \htp\, emission. 

\begin{figure*}[t]
    \centering
    \includegraphics[width=\linewidth]{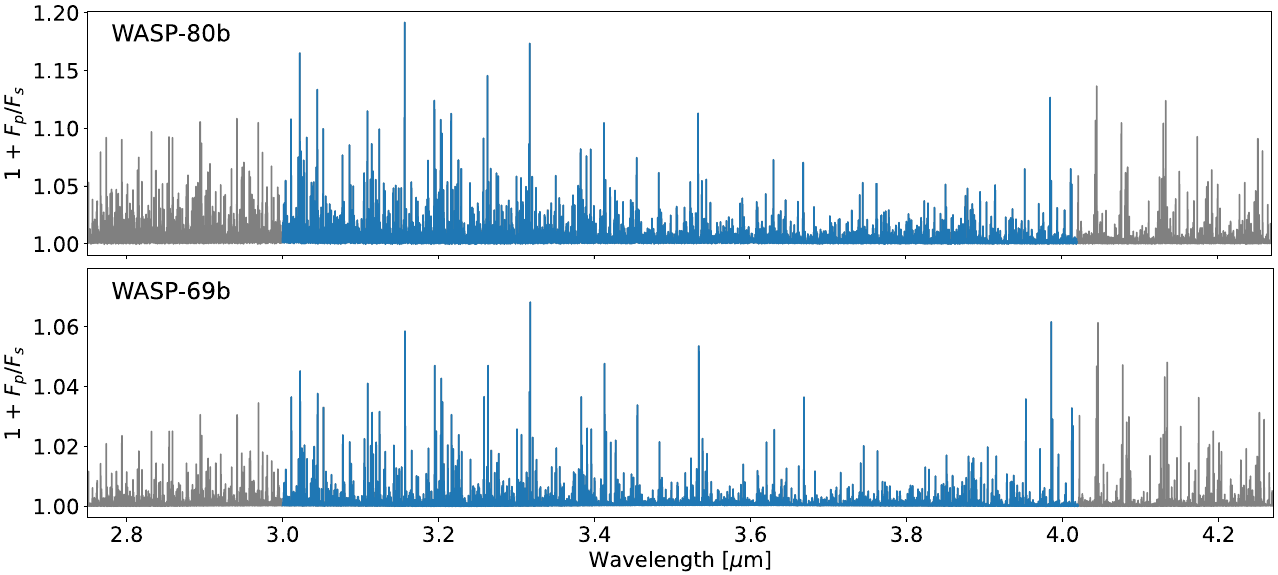}
    \singlespace\caption{An example \htp\, model with log$_{10}$(VMR) = $-4.0$ generated for cross-correlation with the data, as described in Section \ref{sec:models}. The blue section covers the wavelength range of our data. The models are sampled at a resolution of $R \approx 500,000$ but are convolved with NIRSPEC's instrument resolution of $R \approx 75,000$. The planetary continuum is subtracted to yield the line contrast relative to the stellar continuum. The flux ratio is then Doppler-shifted and interpolated onto the same wavelength grid as the data before cross-correlation. }
    \label{fig:models}
\end{figure*}

Both targets were observed in the KL band with an echelle angle of 62.18$^{\circ}$ and cross-disperser angle of 33.46$^{\circ}$, covering 3.00 -- 4.02 \um\, over seven orders, 19 -- 25. This configuration covers the Q(1,0) and Q(3,0) emission lines explored by previous studies \citep{Shkolnik2006, Lenz2016, Gibbs2022}, as well as several other strong \htp \, lines (see Figure \ref{fig:models}). We utilized the 0.144$\arcsec$ x 12$\arcsec$ slit to achieve a resolving power of $\approx$75,000. 

We observed each target for 30-second intervals after which the telescope was nodded in an ABBA pattern, such that the sky could be subtracted out of the data. We took calibrations (i.e., flats) at the beginning and end of each half night. 

\begin{deluxetable*}{l l}[t]
\centering
\tablecaption{\normalsize{Combinations of parameters tested for both WASP-80b and WASP-69b}\label{tab:tested}} 
\tablehead{\colhead{Parameter}& \colhead{Values}}
\startdata
Nights &	all; individual nights; only 2021; 211025+220915 \\
Orders &	all; only 19, 23, and 24 \\
Telluric transmittance masking &	$<$20\%; $<$50\%; $<$80\%; $<$95\%; $<$99\% \\
PCA components removed & 4; 6; 10 \\
Models & thermosphere at 20,000K; 15,000K; 3,000K 
\enddata
\end{deluxetable*}

\begin{figure}[t]
    \centering
    \includegraphics[width=\linewidth]{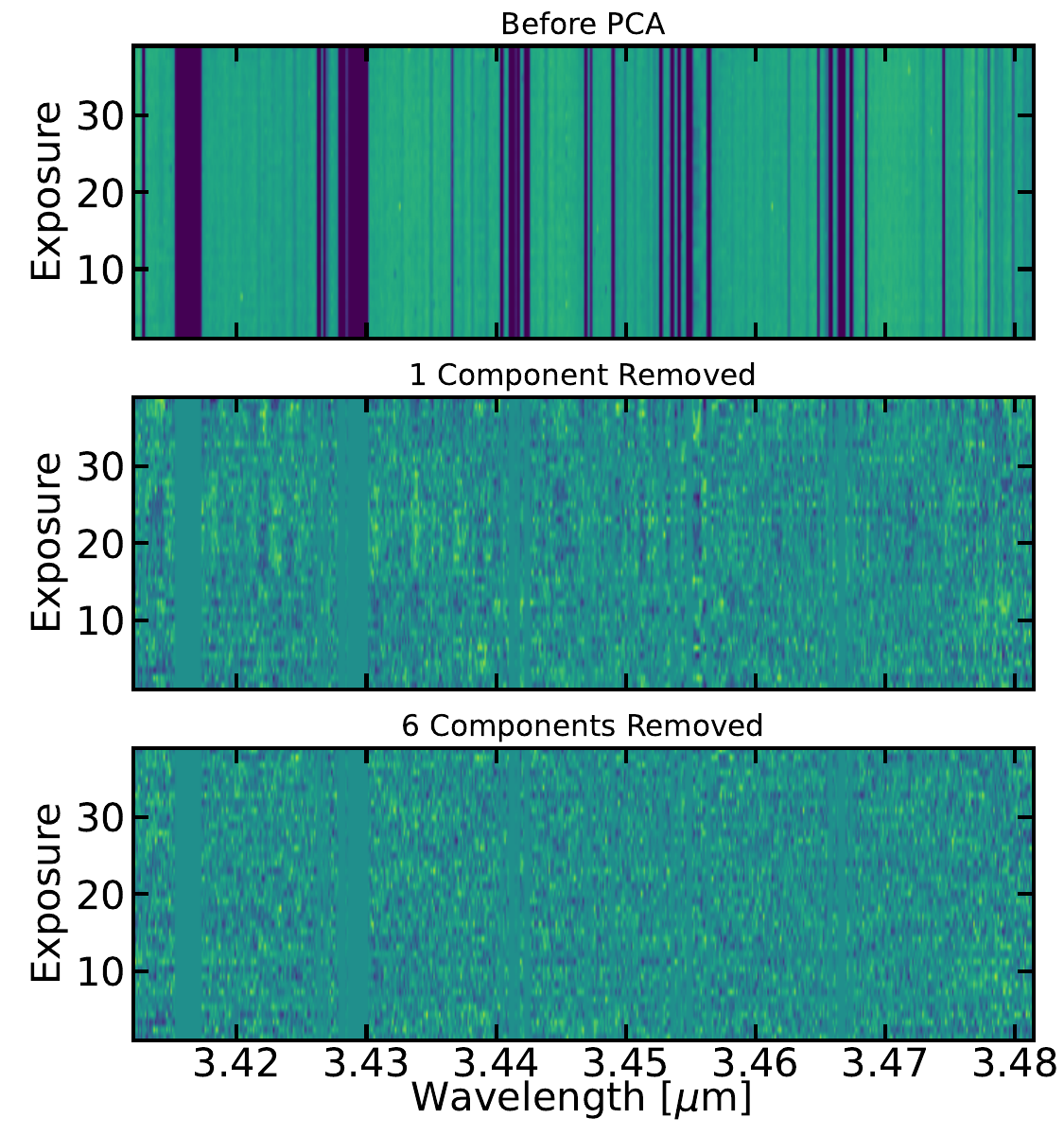}
    \singlespace\caption{PCA-cleaned data from 210913 with different number of principal components removed. The top panel shows the data cube input, where each horizontal cut of the cube is an individual spectrum. The telluric features with transmittance $<$80\% are masked out. The bottom two panels show the remaining spectral cube after removing 1 and 6 principal components, thus removing the telluric and stellar features.}
    \label{fig:steps}
\end{figure}






\subsection{Data Reduction}\label{sec:reduction}



The data were reduced in the following way using Community IRAF v2.17\footnote{\url{https://iraf-community.github.io/}}\citep{IRAF}. The data were first flat-fielded using a dark-subtracted combined flat. A and B nods were subtracted from each other to remove the sky background. Each A-B and B-A spectral pair were then extracted using optimal extraction \citep{OptimalExtraction} and coadded to produce final extracted 1D spectra. Initial data reductions were compared to the Keck data pipeline REDSPEC\footnote{\url{https://www2.keck.hawaii.edu/inst/nirspec/redspec.html}} and no significant differences were identified; therefore, we continued using IRAF for reductions for both targets on all nights. 

Due to the lack of ThAr features at 3 -- 4 \um, rather than wavelength calibrating on arc lines, we instead wavelength calibrated on a telluric template generated from the Planetary Spectrum Generator\footnote{\url{https://psg.gsfc.nasa.gov}} set to the altitude of Mauna Kea and the night of the observing run as there were numerous ($>$10) strong telluric lines in each spectra order. After aligning the spectra, we identified multiple ($>$5) non-saturated telluric features within each order and matched them to the wavelengths in the telluric spectrum. We then fit a third order polynomial to the solution to convert from pixels to wavelength. This produced a successful fit with residuals between the data and telluric model $\leq$ 1 pixel at each order. 

The data were cleaned via sigma-clipping with a sigma of 5 to identify bad pixels. These pixels were then linear interpolated using their nearest two points. We then normalized each set of nightly data by the median of each individual spectrum. Due to the large number of tellurics at these wavelengths, the blaze function was removed by Lowess smoothing \citep{lowess} the data with a fraction of 0.2, after which a fourth-order polynomial was fit to the smoothed data. The original data was divided by this fit to remove the blaze function. This produced a more satisfactory fit than typical convex hull removal used for data from other high-resolution spectrographs such as IGRINS due to the increased number of telluric features at these wavelengths. The resulting data cube can be seen in the top panel of Figure \ref{fig:steps}.

\subsection{Telluric and Stellar Feature Removal}\label{sec:pca}


We utilized principal component analysis (PCA, \citealt{PCA}) to remove the telluric and stellar features in each spectrum. As these features should have negligible motion throughout one night, they can be identified by PCA as something consistent between each spectrum and removed, thus leaving behind the exoplanet signal that will be moving throughout the night as the planet orbits its host star. The principal components were subtracted out rather than normalized to keep the pixels unweighted. 

To prepare for PCA, we created data cubes for each night and masked out any strong telluric features. We tested different masking regimes such as masking features with transmittance $<$20\%, $<$50\%, $<$80\%, $<$95\%, and $<$99\%. All produced similar results; therefore, for convenience and one-to-one comparison, the figures presented in this paper are all with telluric transmittance $<$80\% masked.

We ran the PCA analysis on a night-to-night and order-by-order basis. We tested the cleaning with 1-10 component removal and decided to run the analysis with the removal of 4, 6, and 10 components. The middle and bottom panels of Figure \ref{fig:steps} show the data cube with one and six PCA components removed, respectively. Each component removal does remove some exoplanet signal; therefore, we wanted to be cautious not to remove too many components. While no significant difference in the final results were seen, removal of six components appeared to be the minimum number of components removed with no apparent artifacts in the remaining image. This is additionally in line with previous studies showing the removal of six PCA components to be optimal \citep[e.g.][]{Cabot2019MNRAS.482.4422C, SanchezLopez2019A&A...630A..53S, Pelletier2023Natur.619..491P}. The further figures presented in this paper are therefore all with six PCA components removed. 

The night-to-night resulting PCA-cleaned data cubes were stacked to make a master data cube with all exposures for that target. 

\section{Looking for \htp\, in the Data}\label{sec:search}
\subsection{Direct Measurements}

\begin{figure}[th]
    \centering
    \includegraphics[width=\linewidth]{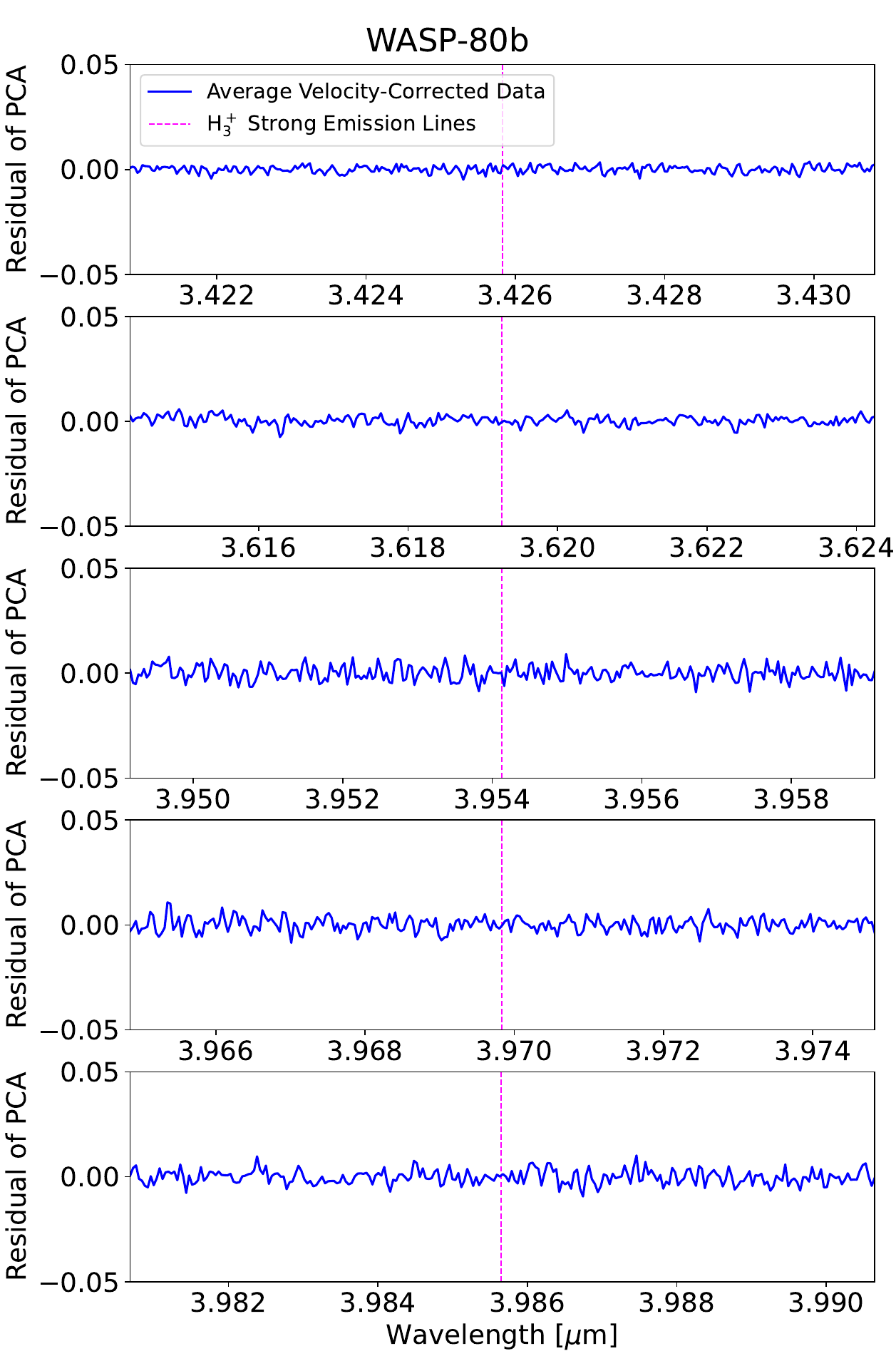}
    \singlespace\caption{Average residuals of the PCA for WASP-80b, i.e. the average spectra with telluric features and stellar spectrum removed. This should just leave behind only the exoplanet spectrum. All spectra have been doppler-shifted to the planet's reference frame before averaging. Each panel is centered on an \htp\, emission line at the transitions R(3,3), R(2,2), Q(1,0), Q(2,1), and Q(3,0). There are no clear \htp\, emission features at these wavelengths.}
    \label{fig:data}
\end{figure}

\begin{figure}[th]
    \centering
    \includegraphics[width=\linewidth]{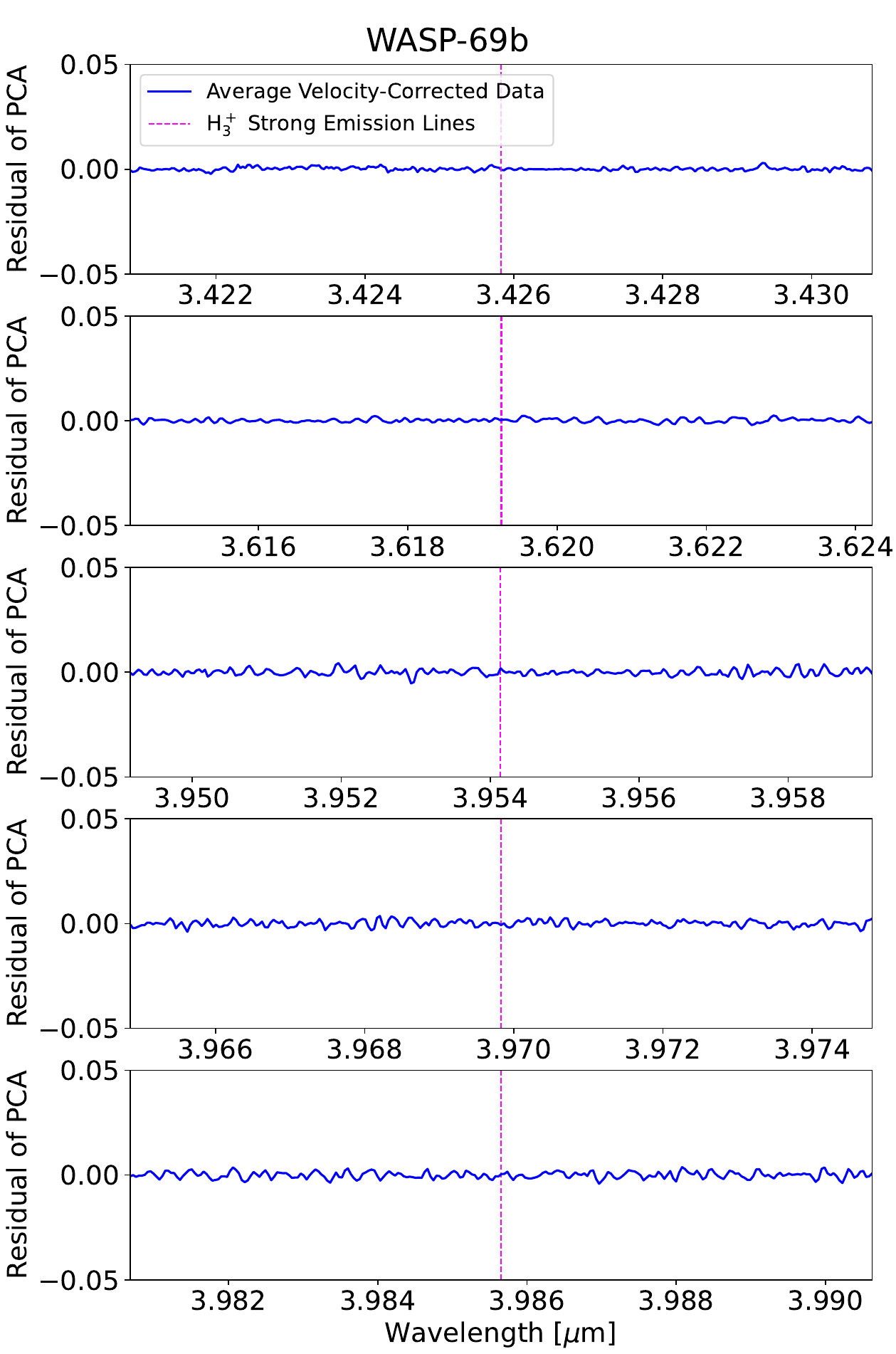}
    \singlespace\caption{Same as Figure \ref{fig:data} but for WASP-69b.}
    \label{fig:ds}
\end{figure}

The most straightforward way to detect \htp\, is to directly search for it in the data. This has been the traditional way to search for \htp\, in planetary and brown dwarf atmospheres \citep{Shkolnik2006, Lenz2016, Gibbs2022, Pineda2024}. We Doppler shifted the PCA-cleaned residual master spectra to the planet's rest frame and took the average of all the spectra. We compared this to the strongest five \htp\, lines covered in our data set (i.e. R(3,3) at rest wavelength 3.42$\mu$m, R(2,2) at 3.62$\mu$m, Q(1,0) at 3.95$\mu$m, Q(2,1) at 3.97$\mu$m, and Q(3,0) at 3.98$\mu$m). Figures \ref{fig:data} and \ref{fig:ds} show the averaged residual spectra and the calculated location where strong \htp\, lines should exist for WASP-80b and WASP-69b, respectively.  We find no indication of planetary emission at the \htp\, lines for neither WASP-80b nor WASP-69b.

\subsection{Cross-Correlation}
We attempt for the first time to search for \htp\, via cross-correlation with a model, a technique which has proven successful for dozens of molecules \citep[e.g.][]{Birkby2013MNRAS.436L..35B, Brogi2019, Pelletier2023Natur.619..491P}. The cross-correlation technique is a method of comparing the data at different points in the planet's orbit with a model of the expected planetary emission. The correlation signal should be the largest at the known system and orbital velocities. The cross-correlation method increases the S/N by $\sqrt{N}$, where $N$ is the number of lines and as we expect potentially dozens of lines with a high enough signal from the \htp\, emission, this method could be advantageous compared to direct measurements.

\begin{figure}[t]
    \centering
    \includegraphics[width=\linewidth]{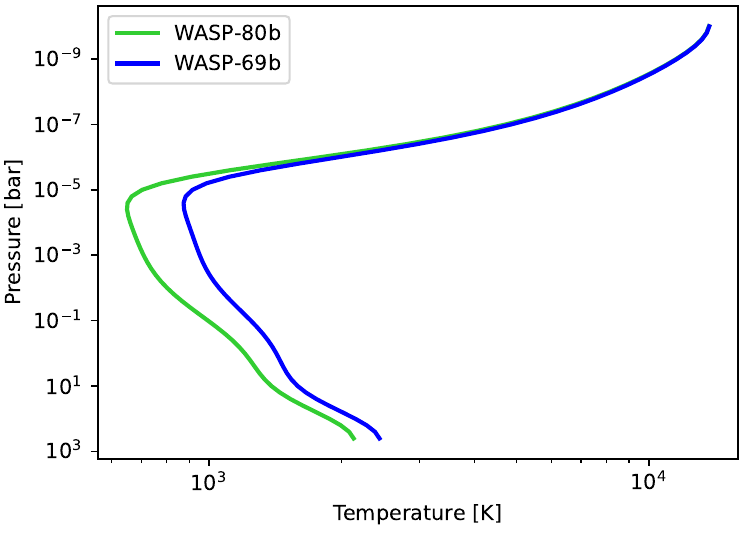}
    \singlespace\caption{Pressure -- temperature profiles used in the generation of the exoplanet models, as described in section \ref{sec:models}.}
    \label{fig:ptprofile}
\end{figure}

\begin{figure*}[h]
    \centering
    \includegraphics[width=\linewidth]{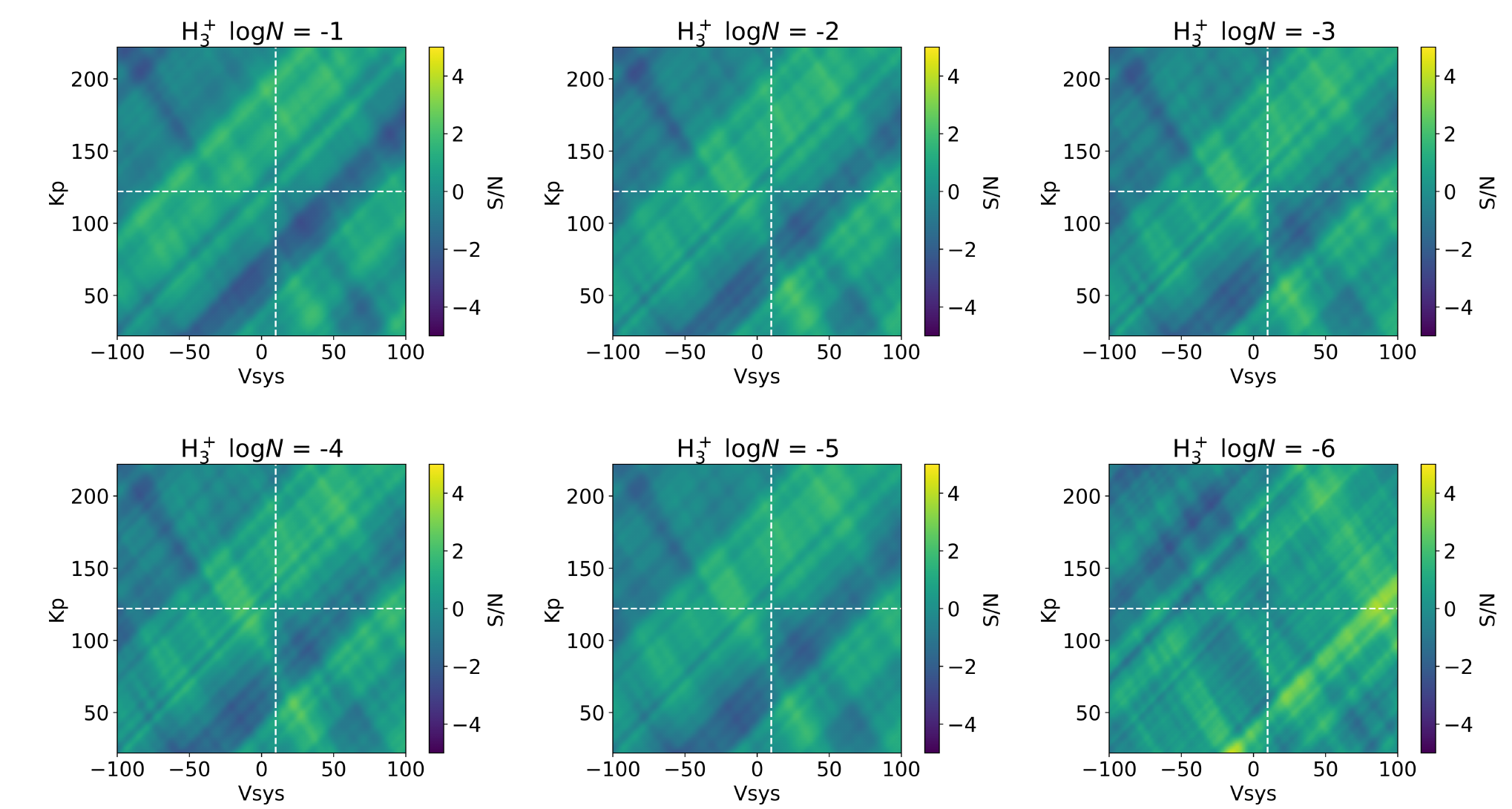}
    \singlespace\caption{Cross-correlation results for WASP-80b for each of the six model abundances. In each case, the dashed white lines represent the expected location of any \htp\, signature. However, no features are identified at a detection threshold of S/N $>$ 3. }
    \label{fig:ccf80}
\end{figure*}

\begin{figure*}[h]
    \centering
    \includegraphics[width=\linewidth]{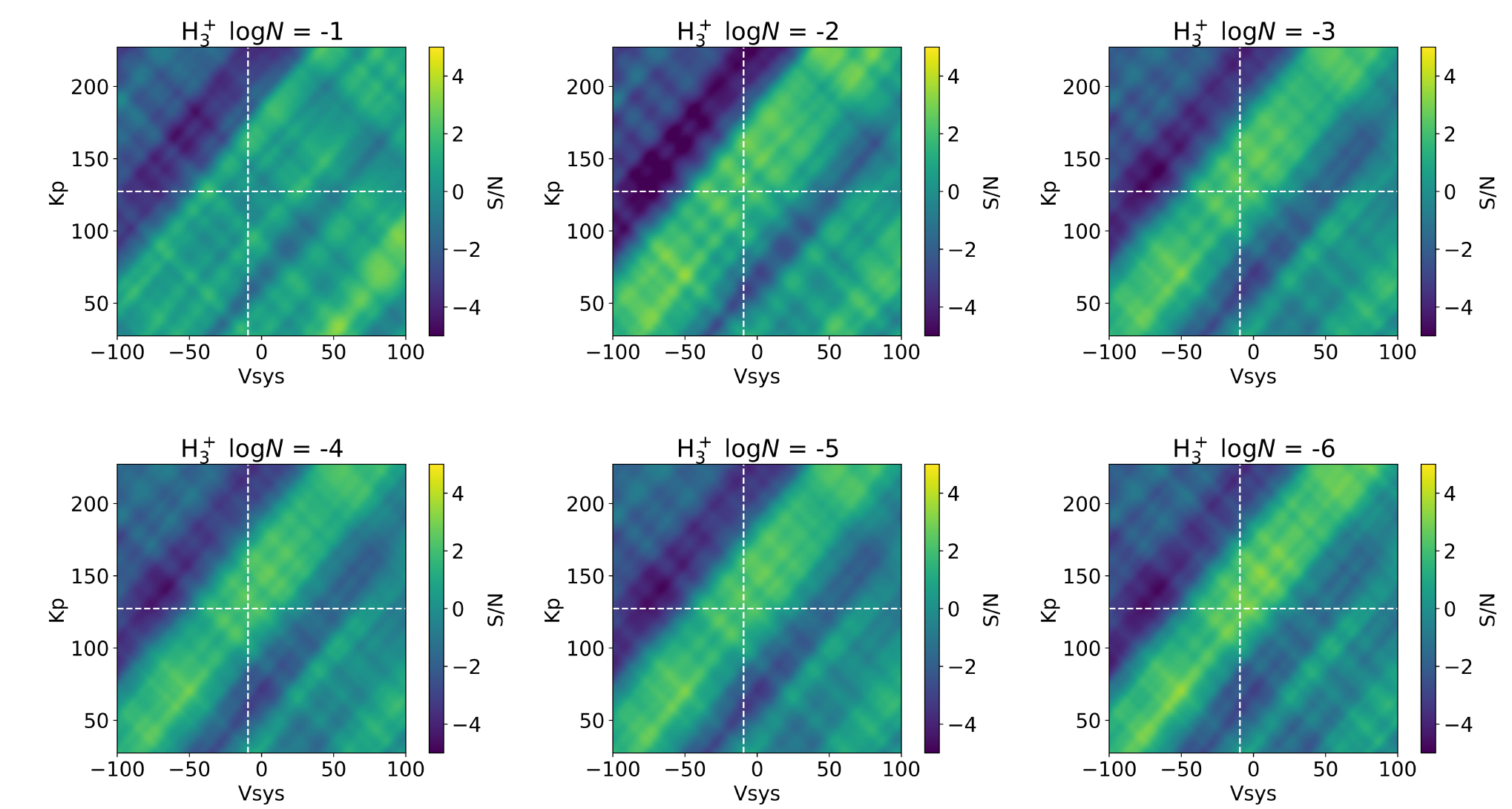}
    \singlespace\caption{Same as Figure \ref{fig:ccf80} but for WASP-69b.}
    \label{fig:ccf69}
\end{figure*}

\begin{figure}[t]
    \centering
    \includegraphics[width=\linewidth]{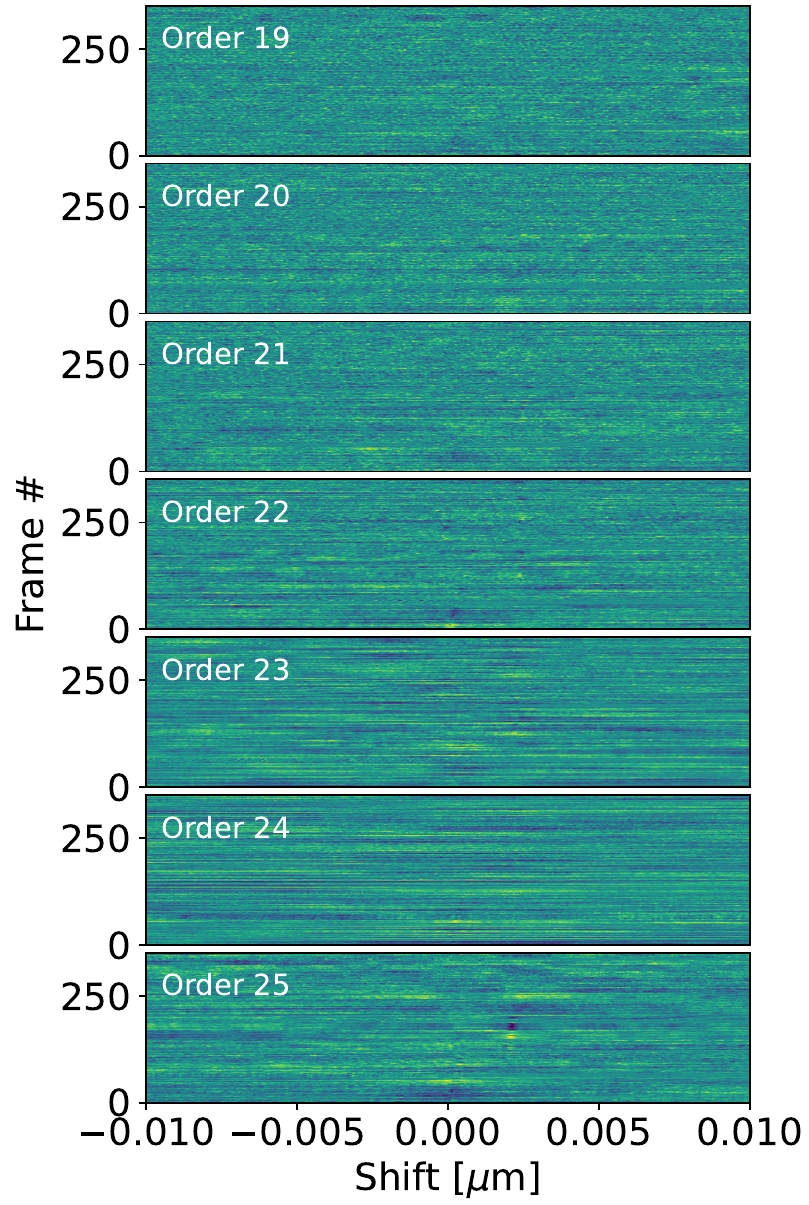}
    \singlespace\caption{Auto-correlation for WASP-80b. The spectra were aligned to the planet's reference frame such that any correlated planet signatures would appear as a peak at a shift of 0 $\micron$. However, no signal is seen in the auto-correlation.}
    \label{fig:ac80}
\end{figure}

\begin{figure}[t]
    \centering
    \includegraphics[width=\linewidth]{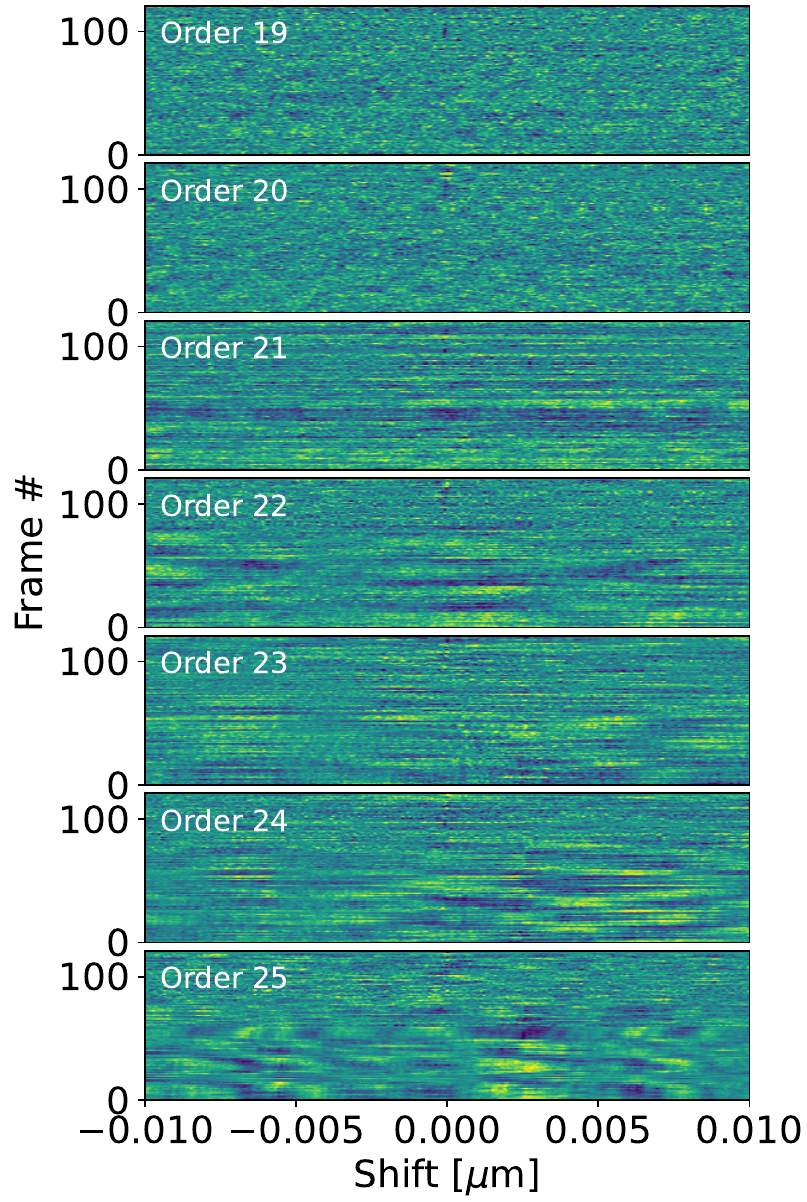}
    \singlespace\caption{Same as Figure \ref{fig:ac80} but for WASP-69b.}
    \label{fig:ac69}
\end{figure}

\begin{figure*}[h]
    \centering
    \includegraphics[width=\linewidth]{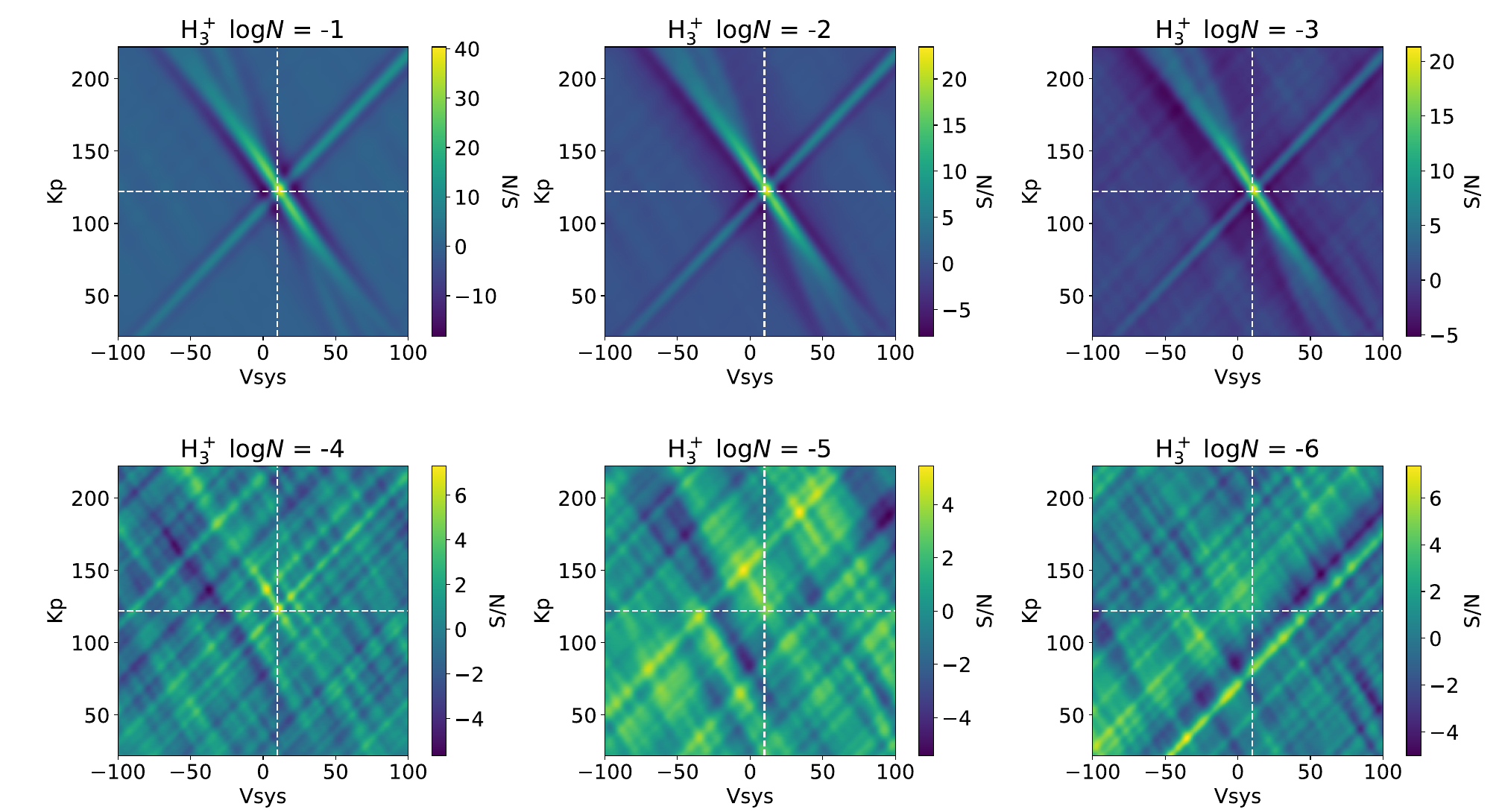}
    \singlespace\caption{Injection-recovery tests for WASP-80b. These results show that we would have recovered \htp\, with strong significance for the abundances -1 $>$ log$N$ $>$ -4. We would not recover -5 $>$ log$N$ $>$ -6 with the current models and data. }
    \label{fig:ccf80-inj}
\end{figure*}

\begin{figure*}[h]
    \centering
    \includegraphics[width=\linewidth]{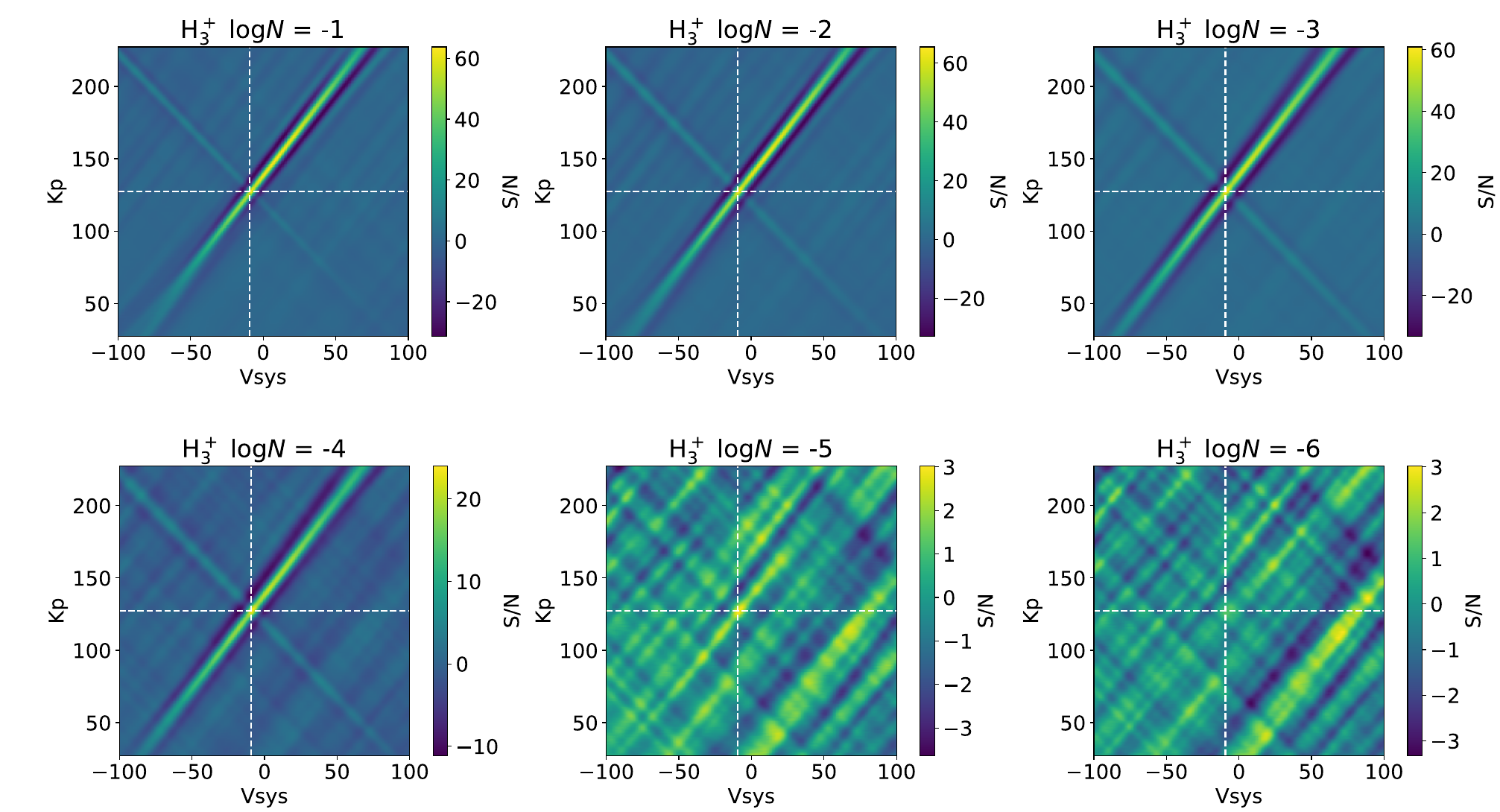}
    \singlespace\caption{Same as Figure \ref{fig:ccf80-inj} but for WASP-69b. In this case, a VMR of log$N$ $=$ -5 would be additionally be recovered with SNR $>$ 3.}
    \label{fig:ccf69-inj}
\end{figure*}

\begin{figure*}[t]
    \centering
    \includegraphics[width=\linewidth]{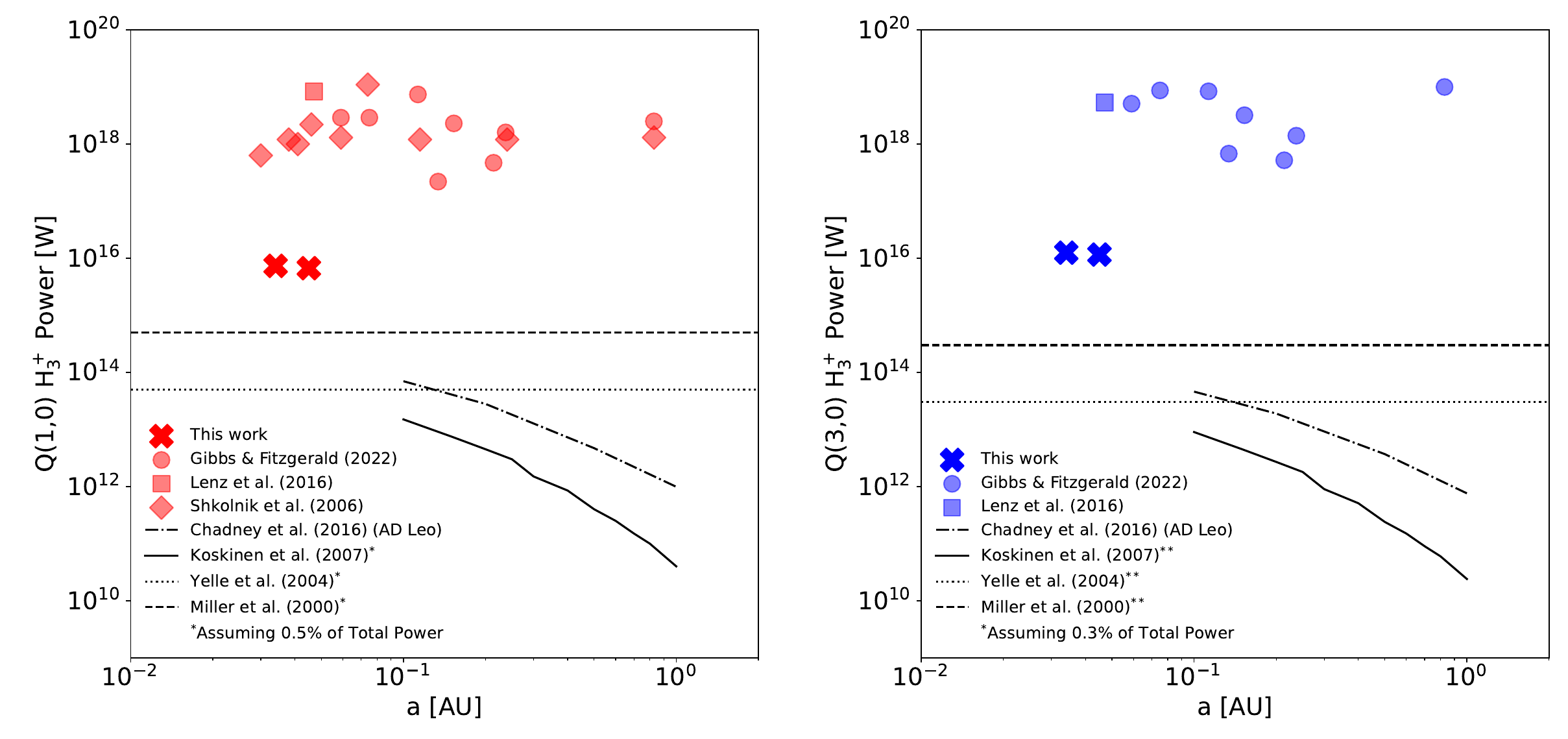}
    \singlespace\caption{\htp\, emission power upper limits in the (left) Q(1,0) and (right) Q(3,0) emission lines. In both figures, the colored points are non-detections, while the lines represent model estimates. Several of the models (\citealt{Miller2000, Yelle2004, Koskinen2007}) are only estimates of total \htp\, power, rather than the Q(1,0) or Q(3,0) individual transition power; therefore, we have assumed that the Q(1,0) transition emits 0.5\% and the Q(3,0) emits 0.3\% of the total \htp\, power. For the model from \citet{Chadney2016A&A...587A..87C}, we use that of AD Leo (M0V). While our upper limits are an order of magnitude below other detection limits in the literature, they are still an order of magnitude away from model estimates of \htp\, emission power for hot Jupiters. However, these models do not include magnetic fields which could elevate the supposed \htp\, emission.}
    \label{fig:upper_limits}
\end{figure*}

\subsubsection{Model Generation}\label{sec:models}
We generated a model planetary spectrum with which to cross-correlate the data. To obtain a reasonable approximation of the vertical thermal structure of each planet's atmosphere, we generated 1D radiative-convective-thermochemical equilibrium (1D-RCTE) models using the ScCHIMERA modeling framework as described in \citet{Arcangeli2018}, \citet{Piskorz2018}, and \citet{Mansfield2022}. For WASP-80b, we assume 5$\times$ solar metallicity and C/O = 0.35, in line with recent JWST observations of WASP-80b \citep{Bell2023}. For WASP-69b, we assume 10$\times$ solar metallicity and C/O = 0.75, in line with JWST observations from \citet{Schlawin2024AJ....168..104S}. We additionally use a heat redistribution efficiency following the predicted trend with a planet equilibrium temperature from \citet{Parmentier2021}. ScCHIMERA outputs dayside averaged 1D pressure-temperature (P-T) and gas volume mixing ratio (VMR) profiles. 

However, the models do not include thermospheres or \htp\, emission. Predictions from \citet{Koskinen2007} estimate that for hot Jupiters, thermospheres can reach temperatures over 20,000K and start to escape hydrodynamically depending on the efficiency of the \htp\, cooling, which can lower thermospheric temperatures down to $\sim$3000K. Because these planets have observed escaping atmospheres, we assume that the temperature must be above 10,000K; therefore, we artificially insert a thermal inversion into this P-T profile beginning at $10^{-5}$ bar and log-linearly increasing up to 15,000 K by $10^{-11}$ bar. We additionally test models with temperatures of 20,000K and 3,000K at the extreme ends of the \htp\, cooling efficiency for thoroughness, although we present 15,000K in this study. To avoid discontinuities, we then smoothed the P-T profile using a Gaussian filter with a standard deviation of 3. The line list used for \htp\, was calculated by \citet{Mizus2017MNRAS.468.1717M} as part of the ExoMol database \citep{exomol}. It utilizes the MARVEL (measured active rotation-vibration energy levels procedure \citep{Furtenbacher2007JMoSp.245..115F, Furtenbacher2012JQSRT.113..929F}, ensuring a highly accurate, empirically driven \htp\, line list.
 
We passed these adjusted P-T profiles through the chemical equilibrium code GGChem \citep{Woitke2018} in order to account for the effects of molecular thermal dissociation in the inversion layer. This provides us with the pressure-dependent VMRs of H$_2$, He, H$_2$O, and CH$_4$. To create such a profile for \htp, we set its deep abundance to zero, and then assume any available H due to the dissociation of H$_2$, H$_2$O, and CH$_4$ is converted to \htp\, up to a set maximum VMR. For both planets, we calculate six possible \htp\, VMR profiles based on this maximum thermospheric abundance, which ranges from $10^{-6}$ to $10^{-1}$ in steps of 1 dex.

We calculate high resolution ($R=500,000$) emission spectra using the adjusted P-T profile and these VMR profiles using a GPU-accelerated version of CHIMERA \citep{Line2013, Brogi2019}. These spectra are then convolved with a Gaussian kernel at NIRSPEC’s nominal spectral resolving power of $R \approx 75,000$ in order to imitate the average instrumental line profile. To convert to the planet-to-star flux ratio $F_p$/$F_s$, we divide the planet spectra by a PHOENIX library stellar spectrum \citep{Husser2013} interpolated at each stellar host's effective temperature and surface gravity. We finally subtract the planetary continuum to yield a planetary line contrast relative to the stellar spectrum as has been successfully carried out for emission cross-correlation by \citet{Herman2022AJ....163..248H}. We individually attempt both using an iterative polynomial fit that removes values greater than the fit until it approaches the bottom edge of the spectrum as well as a high-pass Butterworth filter to remove the planetary continuum, with both yielding negligible differences for the results of the analysis. 


\subsubsection{Planet Signal Search}
Using the PCA-cleaned master data set for each target, we cross correlated each telluric- and stellar-removed residual spectra with the model described above. We utilized an RV lag model, where we test systemic velocities from $-100 \leq V_{sys} \leq 100$ and planetary orbital radial velocity amplitudes from $K_{p, true}-100\, $km/s\,$ \leq K_p \leq K_{p, true}+100\, $km/s, where $K_{p, true}$ is the known planetary orbital amplitude of the system. We Doppler shifted the planet model to each $K_p, V_{sys}$, and observational barycentric velocity $V_{bary}$ using the total velocity $V(t)$ defined as

\begin{equation}
    V(t) = K_p \textnormal{sin}[2\pi\phi(t)] + V_{bary}(t) + V_{sys} 
\end{equation}

\noindent where $\phi$ is the orbital phase of the planet (with 0 defined as the transit midpoint) and the $K_p, V_{sys}$ values are those listed for each planet in Section \ref{sec:system}. For simplicity in calculating the phases, we assume that the eccentricity is 0, since it is very small for both planets: $0.0020^{+0.0100}_{-0.0020}$ \citep{Kokori2023ApJS..265....4K} and 0.0 \citep{Stassun2017AJ....153..136S} for WASP-80b and WASP-69b, respectively. We calculate the correlation coefficient at each of these velocities and then sum the results over each phase to get the $K_p, V_{sys}$ cross-correlation maps. The final S/N reported are these coefficients normalized by the standard deviation of the edges of maps (i.e. the background away from the location of the potential planetary signal).

The resulting $K_p, V_{sys}$ cross-correlation maps can be seen in Figure \ref{fig:ccf80} for WASP-80b and Figure \ref{fig:ccf69} for WASP-69b for each of the six models of \htp\, abundance tested. We additionally ran these analyses with various combinations of nights of data, orders, telluric transmittance masking, and number of PCA components removed, as summarized in Table \ref{tab:tested}. There is no clear detection in any of the maps.

\subsection{Auto-Correlation}
One of the weaknesses of cross-correlation is that it is model dependent. While there have been several papers about the theory of \htp\, emission in extrasolar planets \citep{Miller2000, Yelle2004, Koskinen2007, Chadney2016A&A...587A..87C}, we have yet no observational evidence outside of the solar system to confirm these theories. Therefore, we also test a model-independent correlation method to search for any significant features in this wavelength range that might warrant deeper investigation.

We test two methods of auto-correlation: 1) cross-correlating each individual residual spectrum with a selected reference residual spectrum, and 2) taking the median residual spectrum from each night and cross-correlating each of those with the others. In each case, we first interpolated the spectra on to a grid of uniform wavelength spacing so the correlation lags could be quantified. The spectra were then Doppler shifted to the planet's rest frame, which should yield a correlation peak around a lag of zero \micron\, for each of the auto-correlation maps in Figures \ref{fig:ac80} and \ref{fig:ac69} if the signal is there. Once again, however, we do not detect any planetary signatures.

\section{Upper Limits on \htp\, Emission}\label{sec:upperlimits}

Since no \htp\, emission was detected, we therefore place upper limits on the emission that we should be able to detect with the quality of our data. The power of the cross-correlation technique, by utilizing multiple lines at once, means that this method will yield deeper limits over the direct measurement of the single emission line. Therefore, we focus only on cross-correlation injection-recovery tests for upper limits.



The first step towards understanding the limits of our dataset was first to diagnose whether our observations were sufficient for identifying \htp\, assuming an accurate model. We take the observational data (after blaze- and bad-pixel-correcting but before PCA) and multiply it by the time-dependent (i.e. Doppler shifted) model emission spectrum ($1 + F_p/F_s$) to yield an input model data cube that exactly matches the noise of our data. We run the same analyses including PCA and cross-correlation on this model data. We do this test for each of the six models of different \htp\, abundances. Figures \ref{fig:ccf80-inj} and \ref{fig:ccf69-inj} show the results of these injection-recovery tests. We recover the signal from the abundance models of -1 $>$ log$N$ $>$ -4 with great significance. For WASP-69b, we additionally recover the log$N$ $= -5$ model.  Therefore, we can conclude for these models that either \htp does not appear to be in these planets at these abundances or our model does not match the true physical characteristics of \htp\, emission in hot Jupiters. 

After confirming that our observations have sufficient data to detect \htp\, emission if it were there, we then scale the model emission and rerun the test above to identify the lowest power the \htp\, emission can be before it is undetectable. We utilize a threshold of SNR $>$ 3 as a ``detection". We identify the total power of this emission by removing the model's planetary continuum and calculating the power of all emission lines combined over a range of 2.5 -- 5 $\mu$m. We additionally compute the power of individual emission lines of interest such as the Q(1,0) and Q(3,0) transitions at 3.95$\mu$m and 3.98$\mu$m, respectively. With these tests, we establish a detection limit of our observations that yield an \htp\, total power of $5.32 \times 10^{18}$ W, a Q(1,0) power of $7.35 \times 10^{15}$ W, and Q(3,0) power of $1.25 \times 10^{16}$ W for WASP-80b. For WASP-69b we place an upper limit on \htp\, total power of $1.64 \times 10^{19}$ W, a Q(1,0) power of $6.69 \times 10^{15}$ W, and Q(3,0) power of $1.16 \times 10^{16}$ W.
These results place more stringent limits on hot Jupiter \htp\, emission than previous searches by \citet{Shkolnik2006}, \citet{Lenz2016}, or \citet{Gibbs2022} by at minimum a factor of 30. A comparison of all of the upper limits from this work and the literature can be seen in Figure \ref{fig:upper_limits}.



\section{Conclusions}\label{sec:conclusions}

We carried out a search for auroral \htp\, emission from the hot Jupiters WASP-80b and WASP-69b using high-resolution Keck/NIRSPEC data. We examined the data using three different approaches. The first was a direct search in which we inspected the residual spectra after removing both the telluric features and the stellar spectrum. The second was through cross-correlation with an \htp\, planetary emission model. Finally, we performed an auto-correlation among the data itself, both on an individual and nightly basis. At this time, we find no evidence of \htp\, emission from any of these search methods. Instead, we place upper limits on the total emission of $5.32 \times 10^{18}$ for WASP-80b and $1.64 \times 10^{19}$ for WASP-69b via cross-correlation injection-recovery. 

While these upper limits represent the lowest limits to date on exoplanetary \htp\, emission, these limits still do not yet reach the theoretical limits set by \citet{Miller2000}, \citet{Yelle2004}, \citet{Koskinen2007}, or \citet{Chadney2016A&A...587A..87C}. Our limits are only a factor of 13 larger than the strongest model by \citet{Miller2000}; therefore, it may be possible to reach this regime through longer observations with fuller phase coverage, but will almost certainly be achievable with the next class of Extremely Large Telescopes (ELTs) due to the increased sensitivity and spectral grasp on the planned instruments. It is however possible that these planets may have further inhibited \htp\, emission than expected. \citet{Chadney2016A&A...587A..87C} suggest that while high EUV/XUV radiation does contribute to the production of \htp\, molecules especially in pure H/He atmospheres, there is a limit after which the \htp\, becomes confined at the bottom of the ionosphere where it is more likely to be destroyed by reactions with heavy species. This idea is furthered by \citet{Pineda2024} who suggest that in both brown dwarf and hot Jupiter atmospheres, the \htp\, can undergo reactions with other molecules in the environment such as water or methane on a timescale faster than $\sim$0.01 s emission rate. Future studies with ELTs will have to break this theoretical barrier in order to test whether or not this is the case.

\vspace{0.5cm}
The authors would like to thank the anonymous referee for their time and thoughtful comments. T.R.Y. would like to thank Mike Line and Megan Weiner Mansfield for coding resources; Krishna Kanumalla for helpful discussions; and Matteo Brogi, Jens Hoeijmakers, and Kevin Heng for hosting the ``Exoplanets by the Lake'' workshop in which she learned much about cross-correlation techniques. P.C.B.S. acknowledges support provided by NASA through the NASA FINESST grant 80NSSC22K1598. This research was supported by Keck PI Data Awards \#1665145 and \#1682444. The data presented herein were obtained at Keck Observatory, which is a private 501(c)3 non-profit organization operated as a scientific partnership among the California Institute of Technology, the University of California, and the National Aeronautics and Space Administration. The Observatory was made possible by the generous financial support of the W. M. Keck Foundation. This research has made use of the Keck Observatory Archive (KOA), which is operated by the W. M. Keck Observatory and the NASA Exoplanet Science Institute (NExScI), under contract with the National Aeronautics and Space Administration. This research has made use of the SIMBAD database, operated at CDS, Strasbourg, France; and the European Space Agency (ESA) mission \textit{Gaia} (\url{https://www. cosmos.esa.int/gaia}), processed by the Gaia Data Processing and Analysis Consortium (DPAC, \url{https://www. cosmos.esa.int/web/gaia/dpac/consortium}).

The authors wish to recognize and acknowledge the very significant cultural role and reverence that the summit of Maunakea has always had within the Native Hawaiian community. We are most fortunate to have the opportunity to conduct observations from this mountain. A donation by the authors has been made to the 'Imiloa Astronomy Center to serve the Hawaiian community and help strengthen connections between Native Hawaiian traditions and astronomy.

\software{Astropy \, \citep{astropy:2018},\, Matplotlib \citep{matplotlib},\, Numpy \,\citep{numpy2}, \, Scipy \, \citep{scipy}, \, Spectres \, \citep{spectres}.}

\bibliography{bibliograph}

\end{document}